\documentclass[a4paper,11pt]{article}
\usepackage{pos}

\title{Puzzles in the $B^0_s \rightarrow D_s^{\pm} K^{\mp} $ System}

\author[a,b]{Robert Fleischer}
\author*[a]{Eleftheria Malami}

\affiliation[a]{Nikhef,\\
  Science Park 105, NL-1098 XG Amsterdam, Netherlands}

\affiliation[b]{Department of Physics and Astronomy, Vrije Universiteit Amsterdam,\\
NL-1081 HV Amsterdam, Netherlands}

\emailAdd{Robert.Fleischer@nikhef.nl}
\emailAdd{emalami@nikhef.nl}

\abstract{Non-leptonic $B^0_s \rightarrow D_s^{\pm} K^{\mp}$ transitions are particularly interesting processes to test the Standard Model. As these decays occur via pure tree diagrams, they allow a theoretically clean determination of the angle $\gamma$ of the unitarity triangle. Considering recent LHCb results, an intriguing picture arises, showing tension with the Standard Model. Extracting the branching ratios of the underlying $\bar{B}^0_s \rightarrow D_s^{\pm} K^{\mp}$ modes and combining them with information from semileptonic $B_{(s)}$ decays, we arrive at another puzzling situation, in accordance with similar decays. These patterns could be footprints of New Physics. We present a model-independent strategy to include such New-Physics effects and apply it to the data. Interestingly, new contributions of moderate size could accommodate the data. This formalism offers an exciting probe for new sources of CP violation at the future high-precision B physics era.}

\FullConference{%
  11th International Workshop on the CKM Unitarity Triangle (CKM2021)\\
  22-26 November 2021\\
  The University of Melbourne, Australia
}

\begin{document}
\maketitle
\section{Introduction}
Decays of B mesons are very important to test the flavour sector of the Standard Model (SM) and to explore CP violation. Particularly interesting channels for such studies are the $B^0_s\to D_s^\mp K^\pm$ decays \cite{ADK,RF-BsDsK,DeBFKMST}. In the SM, these modes originate only from tree topologies and offer a theoretically clean determination of the angle $\gamma$ of the Unitarity Triangle (UT). Due to $B^0_s$--$\bar{B}^0_s$ oscillations, interference effects arise between the decay channels $\bar{B}^0_s\to D_s^+K^-$ and $B^0_s\to D_s^+ K^-$, leading to the following time-dependent rate asymmetry for the $f$ final state $D_s^{+} K^-$:
\begin{equation} \label{CP-asym}
	 \mathcal{A}_{CP}(t) =
	  \frac{\Gamma(B^0_s(t)\to f) - \Gamma(\bar{B}^0_s(t)\to f) }
	{\Gamma(B^0_s(t)\to f) + \Gamma(\bar{B}^0_s(t)\to f) }  =
	  \frac{{{C}}\,\cos(\Delta M_s\,t) + {S}\,\sin(\Delta M_s\,t)}
	{\cosh(y_s\,t/\tau_{B_s}) + {\cal A}_{\Delta\Gamma}\,\sinh(y_s\,t/\tau_{B_s})},
\end{equation}
where we introduce the observables $C$, $S$ and ${\cal A}_{\Delta\Gamma}$. A similar relation holds for the CP-conjugate case with the $\overline{C}$, $ \overline{S}$
and $\overline{{\cal A}}_{\Delta\Gamma}$ observables.

An intriguing value of $\gamma = \left( 128 ^{+17}_{-22} \right)^{\circ}$, modulo $180^{\circ }$, which was reported by LHCb \cite{LHCb-BsDsK}, was our motivation to further explore this system. Despite the significant uncertainty, this value is still much larger than the finding of global SM analyses of the UT, which give a value in the regime of $70^{\circ}$ \cite{Amhis:2019ckw,PDG}. A similar range is also found in a recent simultaneous LHCb analysis of various B decays \cite{LHCb:2021dcr}. However, these decays have different dynamics and are characterised by different interference effects. We are interested in shedding more light on the question: could this unexpectedly large value of $\gamma$ imply new sources of CP violation from physics beyond the SM?

\section{Determining the angle $\gamma$}
\begin{figure}[t!]
	\centering
\includegraphics[width = 0.4\linewidth]{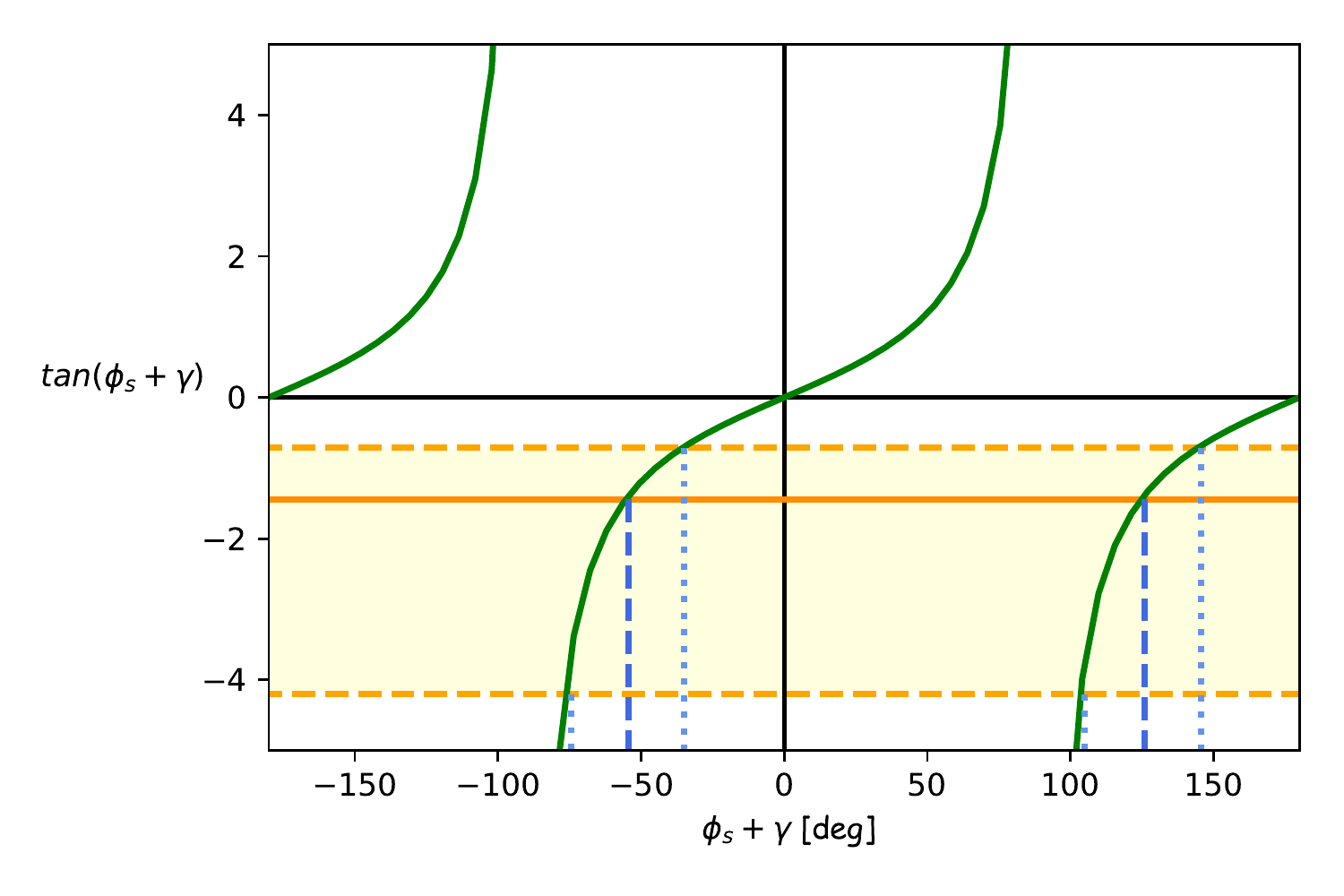} \hspace{0.6cm}
\includegraphics[width = 0.405\linewidth]{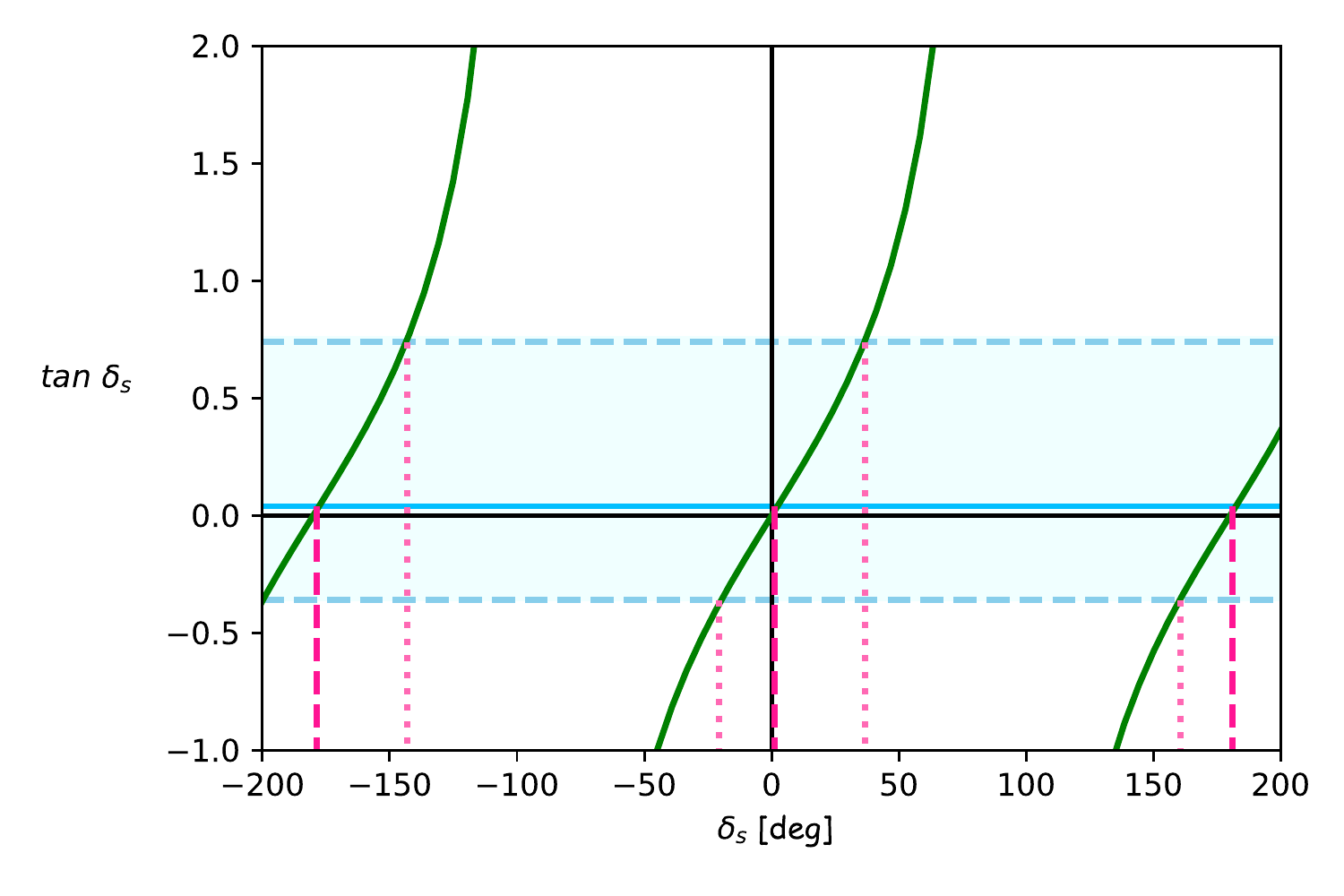}
\vspace*{-0.3truecm}
	\caption{Illustration of $\tan(\phi_s+\gamma)$ (left) and $\tan\delta_s$ (right), where the horizontal lines correspond to Eq. \ref{tan}.}
	\label{fig:tan}
\end{figure}

Let us have a closer look at $\gamma$. We introduce the parameter $\xi$ \cite{RF-BsDsK}: 
\begin{equation}
\xi \propto - e^{-i \phi_s}  [{A(\overline{B}^0_s 
    \rightarrow D_s^{+} K^{-})}/{A(B^0_s \rightarrow D_s^{+} K^{-})}]
    \end{equation}
and similarly $\bar{\xi}$ for the $\bar{f}$ final state $K^+ D_s^{-}$. These quantities are physical observables that measure the strength of the interference effects. The phase $\phi_s$ is the CP-violating $B^0_s$--$\bar B^0_s$ mixing phase and can be determined through $B^0_s \to J/\psi \phi$ and similar modes.  With the help of the measured values of the observables introduced in Eq.\ \ref{CP-asym}, for which:
\begin{equation}
 C=({1-|\xi|^2})/({1+|\xi|^2}),  \quad S= ({2\,\text{Im}{\,\xi}})/({1 + |\xi|^2}), \quad 
 \mathcal{A}_{\Delta \Gamma}=({2\,\text{Re}\,\xi})/({1+|\xi|^2}),
\end{equation}
we can pin down $\xi$ (and similarly $\bar{\xi}$) unambiguously. Assuming the SM expressions for the decay amplitudes leads to the theoretically clean relation 
\begin{equation} \label{multxi}
 {\xi} \times \bar{\xi}= e^{-i2( \phi_s + \gamma)}\ ,
\end{equation}
which allows a clean extraction of  $ \phi_s + \gamma$. LHCb performed a sophisticated fit to their data, assuming $C+\overline{C}=0$, which holds in SM, and obtained $\gamma = \left( 128 ^{+17}_{-22} \right)^{\circ}$. Using $\phi_s=\left(-5^{+1.6}_{-1.5}\right)^\circ$, including penguin corrections  to $B^0_s \to J/\psi \phi$ modes \cite{Barel:2020jvf}, we find the following result \cite{Fleischer:2021cct, Fleischer:2021cwb}:
\begin{equation}\label{gamma-res-1}
\gamma=\left(131^{+17}_{-22}\right)^\circ \ {\text{[modulo $180^\circ$]}}. 
\end{equation}
In view of the tension of this value with the SM, we need to  transparently understand the situation. For this purpose, we utilise the following expressions: 
\begin{equation}
\label{tan}
\tan(\phi_s+\gamma)=-\langle S \rangle_+/ \langle \mathcal{A}_{\Delta \Gamma} \rangle_+=-1.45^{+0.73}_{-2.76},
 \quad \tan\delta_s=\langle S \rangle_- /\langle \mathcal{A}_{\Delta \Gamma} \rangle_+=0.04^{+0.70}_{-0.40} \ 
\end{equation}
with $\langle S \rangle_\pm\equiv (\overline{S}\pm S)/2$ and $\langle \mathcal{A}_{\Delta \Gamma} \rangle_+\equiv(\overline{{\cal A}}_{\Delta\Gamma}+{\cal A}_{\Delta\Gamma})/2$, allowing us to determine  $\phi_s + \gamma$ and $\delta_s$  as shown in Fig.\ref{fig:tan}. We find excellent agreement between this transparent picture and the LHCb fit, confirming the intriguing picture.

\section{Branching ratio information}
The CP-violating observables can be complemented with information from branching ratios, which gives rise to a second puzzling case. We determine the individual branching ratios for the different decay channels. Assuming that $C=-\overline{C}$, in agreement with the LHCb assumption, we obtain the relation between the theoretical and experimental branching ratios \cite{DeBFKMST}:
\begin{equation}\label{SM-BR-3}
 \mathcal{B}_{\text{th}} =  \bar{\mathcal{B}}_{\text{th}}=
\left[\frac{1-y_s^2}{1+y_s\langle {\cal A}_{\Delta\Gamma}\rangle_+}\right]\langle\mathcal{B}_{\text{exp}}\rangle, \ \ {\text{where}} \ \  \langle\mathcal{B}_{\text{exp}}\rangle\equiv \frac{1}{2}\left(\mathcal{B}_{\text{exp}} + \bar{\mathcal{B}}_{\text{exp}}\right)=
\frac{1}{2} \, \mathcal{B}^{\text{exp}}_\Sigma.
\end{equation}
Here we use the average $ \mathcal{B}^{\text{exp}}_\Sigma = (2.27 \pm 0.19) \times 10^{-4}$ \cite{PDG} while $\langle \mathcal{A}_{\Delta \Gamma} \rangle_\pm=0.35 \pm 0.23$ \cite{Fleischer:2021cct, Fleischer:2021cwb} and $y_s \equiv \Delta \Gamma_s / (2 \Gamma_s)=0.062 \pm 0.004$ \cite{PDG}.
We determine $\mathcal{B}_{\text{th}}$ and finally obtain \cite{Fleischer:2021cct, Fleischer:2021cwb}: 
\begin{align}\label{BRbar-Ds+K-}
\mathcal{B}(\bar B^0_s\to D_s^+K^-)_{\text{th}} &=2 \left[{|\xi|^2}/{\left(1+|\xi|^2 \right)} \right]\mathcal{B}_{\text{th}} = (1.94 \pm 0.21) \times 10^{-4}, \\
\mathcal{B}(B^0_s\to D_s^+K^-)_{\text{th}} &=2 \left[{1}/{\left(1+|\xi|^2 \right)} \right]\mathcal{B}_{\text{th}} =(0.26 \pm 0.12) \times 10^{-4}.
\end{align}

For the theoretical SM interpretation, these branching ratios are converted into quantities $|a_1|$, which are phenomenological colour factors that characterise colour-allowed tree decays. Our goal is to extract these parameters in a way that minimises the dependence on uncertainties from CKM parameters and hadronic form factors. The $\bar{B}^0_s \rightarrow D_s^+ K^-$ channel is a prime example where QCD factorisation \cite{Beneke:2000ry} is expected to work excellently. We express the factorised amplitude in terms of CKM matrix elements, the kaon decay constant, the corresponding hadronic form factor and the parameter
\begin{equation}\label{a-eff-1-DsK}
a_{\rm 1 \, eff }^{D_s K}=a_{1}^{D_s K} \left(1+{E_{D_s K}}/{T_{D_s K}}\right),
\end{equation}
where the $a_{1}^{D_s K}$ factor characterises non-factorisable effects entering the colour-allowed tree amplitude $T_{D_s K}$, while $E_{D_s K}$ describes non-factorisable exchange topologies. In the QCD factorization approach, the current state-of-the-art values are $|a_1| \approx 1.07$  \cite{Bordone:2020gao,Huber:2016xod}, with uncertainties at the percent level. For our system, the theoretical values of these parameters are $|a_1^{D_sK}| = 1.07\pm0.02$ and $|a_1^{K D_s}| = 1.1 \pm 0.1$ . We note that additional contributions from exchange topologies, which are non-factorisable, play a minor role and do not indicate any anomalous behaviour \cite{Fleischer:2021cct, Fleischer:2021cwb}. 

We can now determine $|a_1|$ in a clean way, utilising information from $B_{(s)}$ semileptonic decays \cite{FST-BR}.  For the $\bar{B}^0_s \rightarrow D_s^{+}K^{-}$ channel, we use the rate of the theoretical branching ratio of this transition with the differential branching ratio of its partner semileptonic decay $\bar{B}^0_s \rightarrow D_s^{+}\ell^{-} \bar{\nu}_{\ell}$:
\begin{equation}
  R_{D_s^{+}K^{-}}\equiv\frac{\mathcal{B}(\bar{B}^0_s \rightarrow D_s^{+}K^{-})_{\rm th}}{{\mathrm{d}\mathcal{B}\left(\bar{B}^0_s \rightarrow D_s^{+}\ell^{-} \bar{\nu}_{\ell} \right)/{\mathrm{d}q^2}}|_{q^2=m_{K}^2}}=6 \pi^2 f_{K}^2 |V_{us}|^2 |a_{\rm 1 \, eff }^{D_s K}|^2  {\Phi_{\text{ph}}} \left[ \frac{{F_0^{B_s \rightarrow D_s}(m_K^2)}}{{F_1^{B_s \rightarrow D_s}(m_K^2)}} \right]^2, 
    \label{semi}
\end{equation}
where the calculable phase-space factor ${\Phi_{\text{ph}}} \approx 1$. We obtain $|a_{\rm 1}^{D_s K}| = 0.82 \pm 0.11$ \cite{Fleischer:2021cct, Fleischer:2021cwb}. Comparing with the theoretical prediction, this value is surprisingly small, deviating at $2.2 \sigma$ level. Similarly, for the $\bar{B}^0_s \rightarrow K^{+}D_s^{-}$ mode, we obtain $ |a_{\rm 1}^{K D_s}| =0.77 \pm 0.19$, showing again tension with the theoretical result.

\begin{figure}[t!]
	\centering
	\includegraphics[width = 0.4\linewidth]{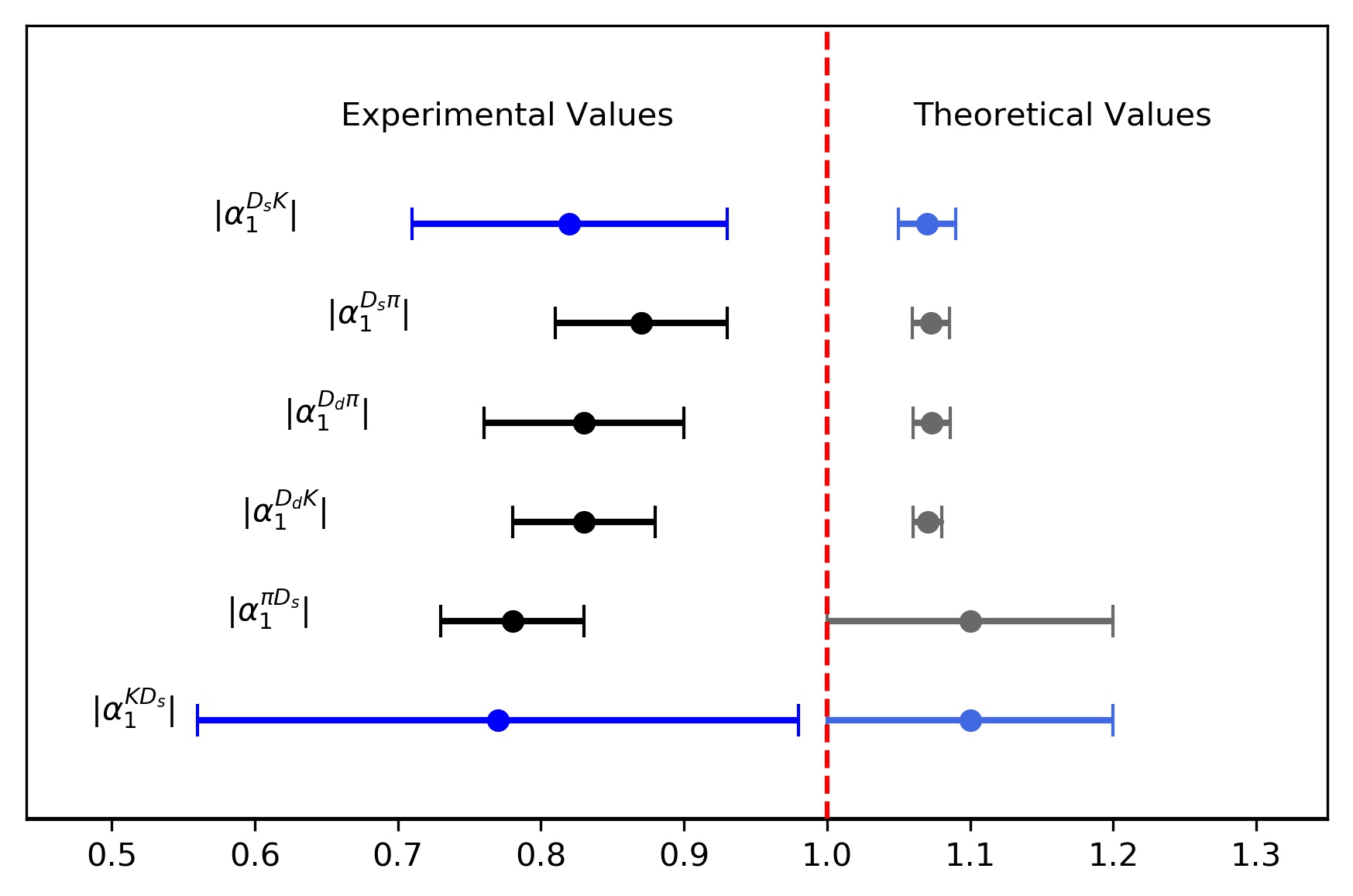}
	\caption{Comparing the $|a_1|$  experimental and theoretical SM values for various decay processes.} \label{fig:aval}
\end{figure}

We complement our analysis, with a detailed look at other $B_{(s)}$ decays with similar dynamics. Interestingly, we observe again a similar pattern, with the parameters following from the data being smaller than the theoretical values, as illustrated in Fig \ref{fig:aval}.  Specifically, we find
\begin{align}
\bar{B}^0_d \rightarrow D_d^+K^- {\text{decay:}} \quad |a_{\rm 1}^{D_d K}|=0.83\pm0.05, \quad {\text{differs at }} 4.8\, \sigma{\text{ level, }}\nonumber \\ 
\bar{B}^0_d\to D_d^+\pi^-  {\text{decay:}} \quad  |a_1^{D_d \pi}|=0.83\pm 0.07, \quad {\text{differs at }} 3.3\,\sigma{\text{ level, }} \nonumber \\ 
\bar{B}^0_s\to D_s^+\pi^-  {\text{decay:}} \quad  |a_1^{D_s\pi}|=0.87\pm0.06, \quad {\text{differs at }} 3.2\,\sigma{\text{ level, }} \nonumber \\ 
\bar{B}^0_d\to \pi^+D_s^-  {\text{decay:}} \quad  |a_{\rm 1}^{\pi D_s}| = 0.78\pm0.05, \quad {\text{differs at }} 2.9\,\sigma{\text{ level. }} \nonumber
\end{align}
This picture complements the intriguing $\gamma$ value. In view of these puzzling cases, we extend our analysis to include New Physics (NP) effects. Studies within physics beyond the SM can be found in \cite{Iguro:2020ndk,Cai:2021mlt,Bordone:2021cca}, and NP effects in non-leptonic tree $B$ decays in \cite{Brod:2014bfa,Lenz:2019lvd}.

\section{Towards New Physics}
\begin{figure}[t!]
	\centering
\hspace{-1cm}\includegraphics[width = 0.41\linewidth]{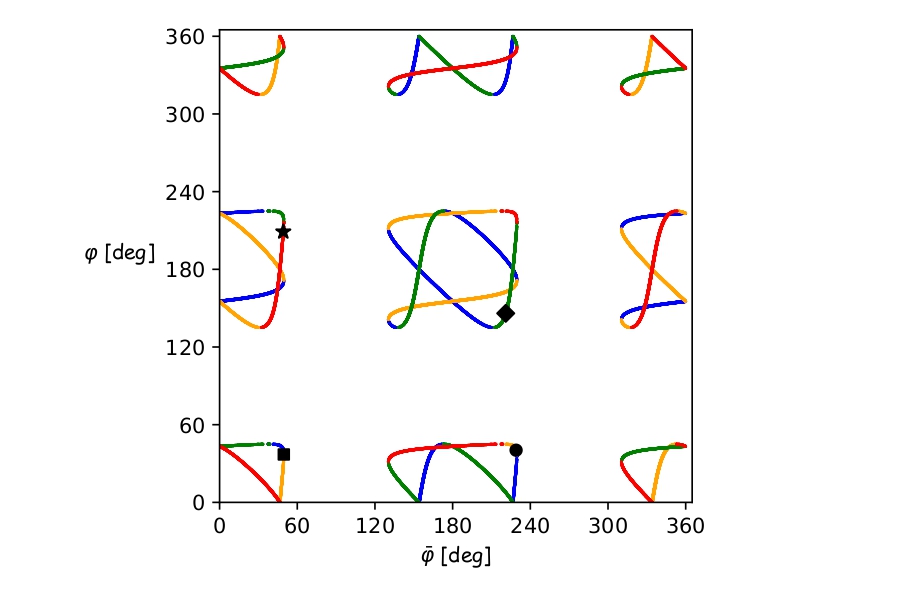}
\hspace{-1.35cm}\includegraphics[width = 0.415\linewidth]{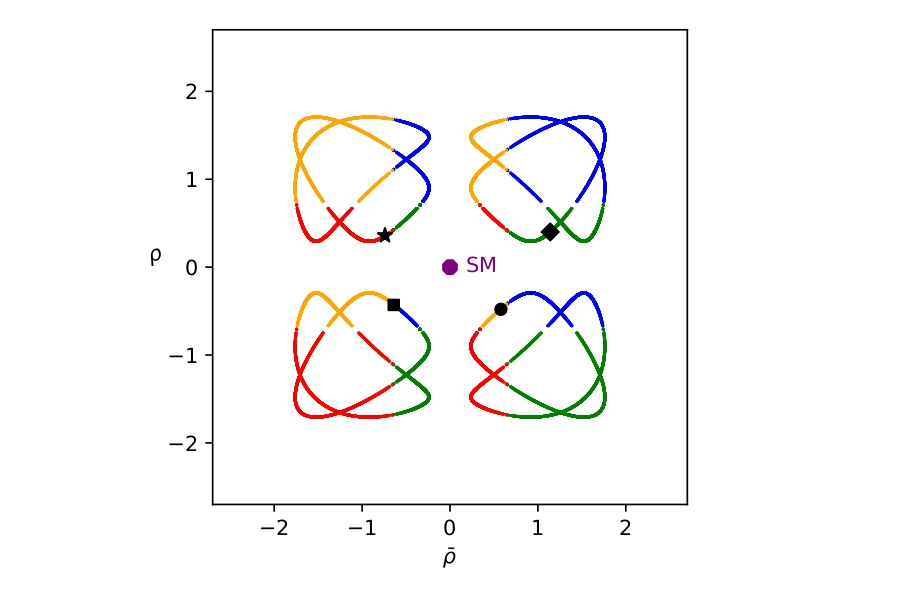}
\hspace{-0.8cm}\includegraphics[width = 0.33\linewidth]{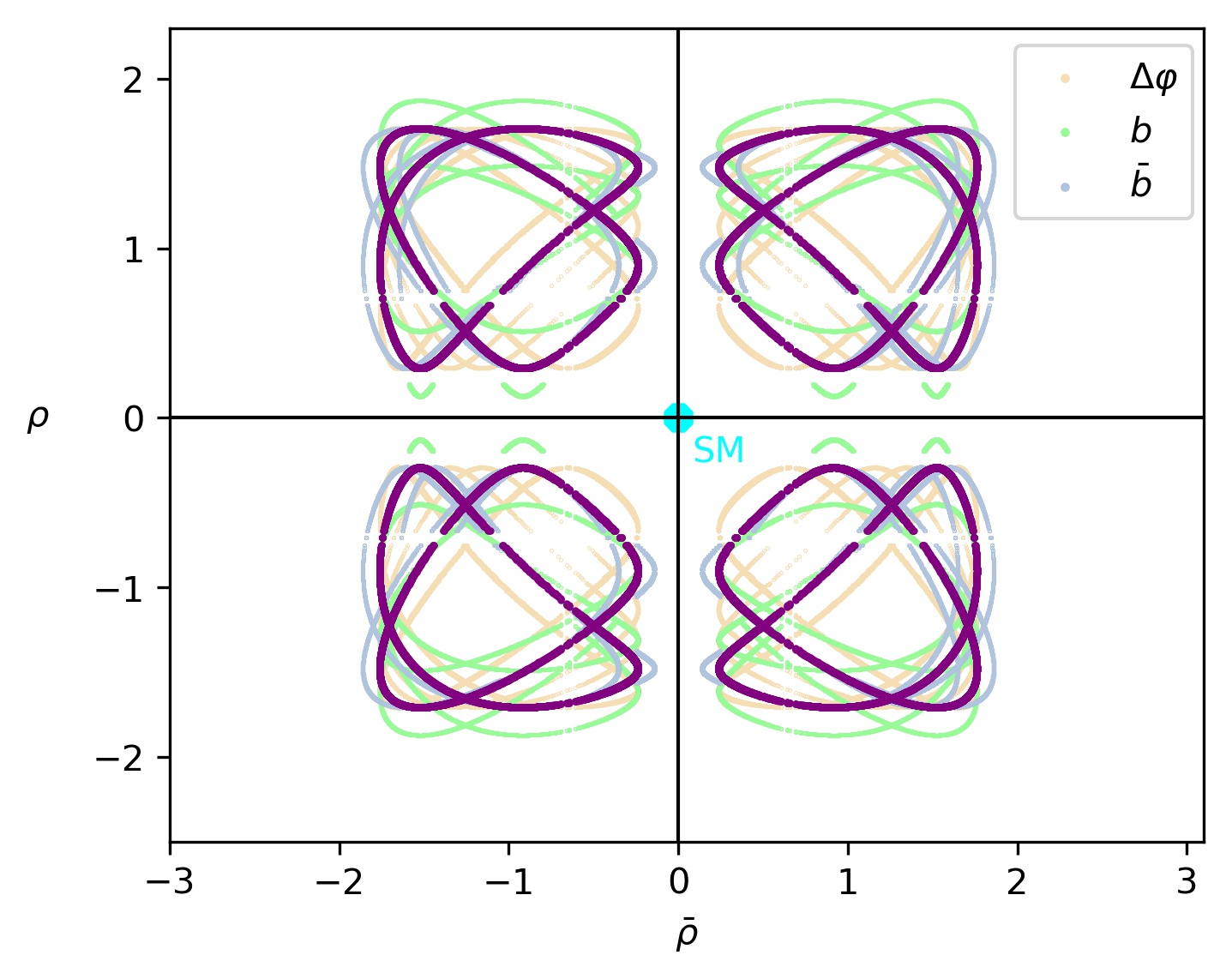}
\vspace*{-0.3truecm}
	\caption{NP correlations in the $\bar{\varphi}$--$\varphi$ plane (left) and the $\bar{\rho}$--$\rho$ plane (center). We pick some points from the $\bar\varphi$--$\varphi$ plane and show the corresponding values in the $\bar\rho$--$\rho$ plane. The different colours come from different sign combinations of the $\rho \bar{\rho}$ term in the $\tan{\Delta \phi}$ expression. Uncertainties in the $\bar{\rho}$--$\rho$ plane (right).}
	\label{fig:centrrhophi-1}
\end{figure}

If these puzzles are due to NP, it would have to enter at the amplitude level, as NP in mixing is included through the use of $\phi_s$ from data. Introducing the NP parameters
\begin{equation}
\bar{\rho} \, e^{i \bar{\delta}} e^{i \bar{\varphi}}  \equiv { A(\bar{B}^0_s \rightarrow D_s^+ K^-)_{{\text{NP}}} }/{  A(\bar{B}^0_s \rightarrow D_s^+ K^-)_{{\text{SM}}} }, 
\end{equation}
with $\bar{\delta}$ and $\bar{\varphi}$ denoting the strong and weak NP phases, respectively (and similarly, ${\rho},{\delta},{\varphi}$ for the CP-conjugate case), we parametrise the amplitudes.  Then, we can generalise the expression of the observables $\xi$ and $\bar{\xi}$ as well as the assumption $ C+\overline{C}=0$ which was used by the LHCb Collaboration. The product can be generalised as follows \cite{Fleischer:2021cct, Fleischer:2021cwb}:
\begin{equation}
\xi \times \bar{\xi}  = \sqrt{1-2\left[\frac{C+\bar{C}}{\left(1+C\right)\left(1+\bar{C}\right)}
\right]}e^{-i\left[2 (\phi_s +\gamma_{\rm eff})\right]}, \quad  {\text{with}} \quad \gamma_{\rm eff}\equiv \gamma-\frac{1}{2}\left(\Delta\varphi+\Delta\bar{\varphi}\right).
\label{eq:generxi}
\end{equation}
Here $\gamma$ enters with a shift due to the CP-violating NP phases, resulting in an effective angle $\gamma_{\rm eff}$. Setting the strong phases $\delta$ and $\bar{\delta}$ to 0, in agreement with factorization, we find
\begin{equation}
  \tan{\Delta \phi}=\frac{\rho \sin{\phi} + \bar{\rho} \sin{\bar{\phi}} + \rho \bar{\rho} \sin{(\bar{\phi} + \phi )}}{1 + \rho \cos{\phi} + \bar{\rho} \cos{\bar{\phi}} + \rho \bar{\rho} \cos{(\bar{\phi} + \phi)}}.
  \label{eq:tan}
\end{equation}
Using $\gamma=(70 \pm 7)^\circ$, we obtain the numerical value $\Delta\varphi=-(61 \pm 20)^\circ$.

In the presence of NP, we use CP-averaged ratios $\langle R \rangle$, and introduce
\begin{align}
    b&= 1 + 2 \rho \cos{\delta} \cos{\phi} + \rho^2 = \frac{ \langle R_{D_s^{+}K^{-}} \rangle}{6 \pi^2 f_{K}^2 |V_{us}|^2 |a_{1}^{D_s K}|^2 X_{D_s K}}=0.58 \pm 0.16,\\
     {\bar{b}}&= 1 + 2 {\bar{\rho}} \cos{{\bar{\delta}}} \cos{{\bar{\phi}}} + {\bar{\rho}}^2 = \frac{  \langle R_{K^{+}D_s^{-}} \rangle}{6 \pi^2 f_{D_s}^2 |V_{cs}|^2 |a_{1}^{K D_s}|^2 X_{K D_s}}=0.50 \pm 0.26,
     \label{betabar}
\end{align}
allowing us to probe the NP parameters. In SM, these quantities are equal to 1. 

We would like to constrain the NP parameters through the data. Employing $b$ and $\bar{b}$, we may express $\rho$ and $\bar{\rho}$ in terms of $\varphi$ and $\bar{\varphi}$, respectively. Therefore, Eq.~(\ref{eq:tan}) allows us to calculate a correlation between ${\varphi}$ and $\bar\varphi$ (Fig.~\ref{fig:centrrhophi-1}, left).  Finally, using $\bar{\rho}(\bar\varphi)$ and ${\rho}(\varphi)$, we may obtain the values of $\bar{\rho}$ and ${\rho}$ for a given value of $\bar\varphi$ and $\varphi$, respectively, thereby determining a correlation in the $\bar{\rho}$--$\rho$ plane (Fig.~\ref{fig:centrrhophi-1}, central). We note that the SM point corresponding to the origin $(0,0)$ in the $\bar{\rho}$--$\rho$ plane is excluded. Values, in the regime around 0.5, could accommodate the central values of the current data, therefore resolving the puzzling patterns in the CP violation measurements as well as in the branching ratios. In the right plot of Fig.~\ref{fig:centrrhophi-1}, we show the impact of the uncertainties of the input quantities $\bar{b}$, $b$  and $\Delta \varphi$, varying each one of them separately (lighter colours). We now observe that NP could accommodate the data with NP contributions as small as about $30\%$ of the SM amplitudes.  

\section{Conclusions}
The $B^0_s\to D_s^\mp K^\pm$ system shows two puzzles: the CP violation measurements, reflected by the value of $\gamma$, and the branching ratios of the individual channels. We extract the parameters $a_1$ from the data in a clean way and we find consistent patterns in $B_{(s)}$ decays with similar dynamics. In view of these puzzles, we have developed a model-independent strategy, generalising the analysis of the $B^0_s\to D_s^\mp K^\pm$ system in order to include effects from physics beyond the SM. Applying our formalism to the current data, we calculate correlations between the NP parameters and their CP-violating phases. Interestingly, we observe that both the CP violation and the branching ratio measurements can be described with NP contributions at the level of $30\%$ of the SM amplitudes. In the future high-precision era of $B$ physics, when much more data will be available, this strategy can be fully exploited. It will be exciting to see  whether the tantalising question will be answered: Could new sources of CP violation be established?


\begin{thebibliography}{99}
\bibitem{ADK}
R.~Aleksan, I.~Dunietz and B.~Kayser,
Z. Phys. C \textbf{54} (1992), 653-660


\bibitem{RF-BsDsK}R.~Fleischer,
Nucl. Phys. B \textbf{671} (2003), 459-482

\bibitem{DeBFKMST}
K.~De Bruyn \textit{et al.},
Nucl. Phys. B \textbf{868} (2013), 351-367
  
\bibitem{LHCb-BsDsK}R.~Aaij \textit{et al.} [LHCb],
JHEP \textbf{03} (2018), 059

\bibitem{Amhis:2019ckw}
Y.~S.~Amhis \textit{et al.} [HFLAV],
Eur. Phys. J. C \textbf{81}, no.3, 226 (2021)

\bibitem{PDG}P.~A.~Zyla \textit{et al.} [Particle Data Group],
PTEP \textbf{2020} (2020) no.8, 083C01

\bibitem{LHCb:2021dcr}
R.~Aaij \textit{et al.} [LHCb],
[arXiv:2110.02350 [hep-ex]].

\bibitem{Barel:2020jvf}
M.~Z.~Barel, K.~De Bruyn, R.~Fleischer and E.~Malami,
J. Phys. G \textbf{48} (2021) no.6, 065002

\bibitem{Fleischer:2021cct}
R.~Fleischer and E.~Malami,
[arXiv:2109.04950 [hep-ph]].

\bibitem{Fleischer:2021cwb}
R.~Fleischer and E.~Malami,
[arXiv:2110.04240 [hep-ph]].


\bibitem{Beneke:2000ry}
M.~Beneke, G.~Buchalla, M.~Neubert and C.~T.~Sachrajda,
Nucl. Phys. B \textbf{591}, 313-418 (2000)

\bibitem{Bordone:2020gao}
M.~Bordone \textit{et al.},
Eur. Phys. J. C \textbf{80}, no.10, 951 (2020)

\bibitem{Huber:2016xod}
T.~Huber, S.~Kr\"ankl and X.~Q.~Li,
JHEP \textbf{09}, 112 (2016)

 \bibitem{FST-BR}
R.~Fleischer, N.~Serra and N.~Tuning,
Phys. Rev. D \textbf{83} (2011), 014017

\bibitem{Iguro:2020ndk}
S.~Iguro and T.~Kitahara,
Phys. Rev. D \textbf{102} (2020) no.7, 071701

\bibitem{Cai:2021mlt}
F.~M.~Cai, W.~J.~Deng, X.~Q.~Li and Y.~D.~Yang,
JHEP \textbf{10} (2021), 235

\bibitem{Bordone:2021cca}
M.~Bordone, A.~Greljo and D.~Marzocca,
JHEP \textbf{08} (2021), 036


\bibitem{Brod:2014bfa}
J.~Brod \textit{et al.},
Phys. Rev. D \textbf{92} (2015) no.3, 033002

\bibitem{Lenz:2019lvd}
A.~Lenz and G.~Tetlalmatzi-Xolocotzi,
JHEP \textbf{07} (2020), 177

\end{thebibliography}
\end{document}